\documentclass[aps,prb,twocolumn,superscriptaddress,showpacs]{revtex4}

\usepackage{hyperref,latexsym}
\usepackage{graphicx,color,pstricks}
\usepackage{epsfig,epsf,bm}

\begin{document}
\title{The Universality and stability for a dilute Bose gas with a Feshbach resonance}
\author{Yu-Li Lee}
\email{yllee@cc.ncue.edu.tw} \affiliation{Department of Physics,
National Changhua University of Education, Changhua, Taiwan,
R.O.C.}
\author{Yu-Wen Lee}
\email{ywlee@thu.edu.tw} \affiliation{Department of Physics,
Tunghai University, Taichung, Taiwan, R.O.C.}
\begin{abstract}
 We study the bosonic atoms with a wide Feshbach resonance at zero
 temperature in terms of the renormalization group. We indicate that
 this system will always collapse in the dilute limit. On the side
 with a positive scattering length, the atomic superfluid is an
 unstable local minimum in the dilute limit and it determines the
 thermodynamics of this system within its lifetime. We calculate
 the equilibrium properties at zero temperature in the unitary
 regime. They exhibit universal scaling forms in the dilute limit
 due to the presence of a nontrivial zero temperature, zero density
 fixed point. Moreover, we find that the $T=0$ thermodynamics of
 this system in the unitary limit is exactly identical to the one
 for an ideal Fermi gas.
\end{abstract}


\maketitle

\section{Introduction}

The bosonic atoms with Feshbach resonances have been one of the
exciting ultrocold systems studied so far. On the side with a
negative scattering length, the controlled collapsing and
exploding dynamics has been observed\cite{exp1}. On the other side
where the scattering length becomes positive, molecules can be
produced and this system becomes a mixture of atoms and molecules.
Despite the reduced lifetime of molecules in the unitary regime
due to the enhanced three-body recombination\cite{FKSW}, quite
remarkable progress has been made and the properties of the
strongly repulsive ultracold bosonic atoms have also been studied
experimentally\cite{exp2}.

A mean-field phase diagram for the Bose gas with an $s$-wave
Feshbach resonance has been proposed\cite{RPW}. At zero
temperature, there are two phases: the atomic superfluid (ASF) in
which both atoms and molecules condense and the molecular
superfluid (MSF) in which only the molecules condense. Hence,
there is a quantum phase transition between the ASF and MSF
phases\cite{RPW,LL}. The stability of these phases at $T=0$ was
examined within the mean-field level\cite{BM}. Recently, a
Nozi\`eres-Schmitt-Rink formalism was developed to study the
finite temperature properties of this system\cite{KMDS}. The
ground state of the cold Bose gas with a large positive scattering
length has also been studied by the variational wavefunction
approach\cite{wf,SZ}. Near the Feshbach resonance, various
physical quantities were found to exhibit universal scaling
forms\cite{SZ}.

Although the idea of the Feshbach resonance has its root in the
few-body physics, it was shown that, for a dilute quantum gas, it
is possible to cast the low-energy physics near the Feshbach
resonance and the associated universal properties into the
framework of the renormalization group (RG) due to the presence of
a nontrivial zero-density fixed point\cite{NS}. Bearing this
picture in mind, the low-energy properties of a dilute Fermi gas
with a wide Feshbach resonance in the unitary regime can be
computed either by an $\epsilon$ expansion\cite{NS1} ($\epsilon
=4-d$) or a large N expansion\cite{NS,VSR}. Here we will extend
the idea in Ref. \onlinecite{NS} to the dilute Bose gas with a
wide Feshbach resonance. In contrast to the case of fermionic
atoms, the self-interactions between atoms and molecules must be
included in order to guarantee the stability of the uniform phases
although these operators are irrelevant in the sense of RG. This
point was also emphasized in Ref. \onlinecite{BM}. The resulting
stability conditions impose a lower bound on the value of the
chemical potential for atoms. Hence, we conclude that a
Feshbach-resonant (FR) Bose gas will always collapse in the dilute
limit. Moreover, in the dilute limit, the ASF state can appear in
the guise of an unstable local minimum for the free energy
density, while the MSF state cannot exist. The ASF state will
eventually collapse. However, within its lifetime, the properties
of this system may be captured by the ASF state when we approach
the unitary limit from the side with a positive scattering length.
Then, we use the $\epsilon$ expansion to calculate various
properties of the ASF state at $T=0$, including the relation
between the chemical potential and detuning, the equation of
state, and the sound velocity. We show that the thermodynamical
properties at $T=0$ exhibit universal scaling forms near the
Feshbach resonance due to the presence of a nontrivial
zero-temperature, zero-density fixed point. Surprisingly, we find
that the $T=0$ thermodynamics of the ASF state in the unitary
limit is exactly identical to the one for an ideal Fermi gas. This
correspondence was also noticed in Refs. \onlinecite{wf} and
\onlinecite{SZ} from a different approach. Our main results are
summarized in Figs. \ref{bfp3}, \ref{asfmu3}, \ref{asfp3}, and
\ref{asfv3}.

The rest of this paper is organized as follows: In Sec. \ref{RG},
we present the RG theory to fix the notation. We discuss the
stability of the dilute Bose gas with a wide Feshbach resonance
and its phase diagram at $T=0$ in Sec. \ref{phase}. The
equilibrium properties of the ASF state at $T=0$ are calculated in
Sec. \ref{asf}. The last section is devoted to a conclusive
discussion and comparison with previous works.

\section{The renormalization-group theory}
\label{RG}

We start with the action for the atom-molecule model in the
grand-canonical ensemble \cite{RPW}:
\begin{eqnarray}
 S &=& \! \int^{\beta}_0 \! \! d\tau \int \! \! d^dx\Psi_a^{\dagger} \!
 \left(\partial_{\tau}-\frac{\nabla^2}{2m}-\mu\right) \! \Psi_a
 \nonumber \\
 & & + \! \int^{\beta}_0 \! \! d\tau \int \! \! d^dx\Psi_m^{\dagger} \!
 \left(\partial_{\tau}-\frac{\nabla^2}{2M}-2\mu+\delta_0\right) \! \Psi_m
 \nonumber \\
 & & + \! \int^{\beta}_0 \! \! d\tau \int \! \! d^dx~U \ ,
 \label{feshbact1}
\end{eqnarray}
where
\begin{eqnarray*}
 U &=& g_0\left(\Psi^{\dagger}_m\Psi_a\Psi_a+{\mathrm H.c.}\right)+u_3
 |\Psi_a|^2|\Psi_m|^2 \\
 & & +\frac{u_1}{2}|\Psi_a|^4+\frac{u_2}{2}|\Psi_m|^4 \ .
\end{eqnarray*}
Here $\Psi_a$ and $\Psi_m$ are the annihilation operators of atoms
and molecules, respectively. $m$ and $M$ are the masses of the
atom and the molecule, respectively. Galilean invariance requires
that $M=2m$. We will see later that $\delta_0=0$ does not
correspond to the unitary limit ($a_s^{-1}\rightarrow 0$ where
$a_s$ is the scattering length between atoms) due to
renormalization. Hence, we define
\begin{eqnarray*}
 \delta =\delta_0-\delta_c \ ,
\end{eqnarray*}
where $\delta_c$ is a counterterm which is chosen such that
$\delta =0$ just corresponds to the unitary limit. We further
define $\mu_m=2\mu -\delta$.

We see that the action $S$ [Eq. (\ref{feshbact1})] has a quantum
critical point (QCP) at $\mu =0=\delta$ and $T=1/\beta =0$, which
will be dubbed as the Gaussian fixed point. That is, $S$ is
invariant against the scaling transformation
\begin{equation}
 \bm{x}\rightarrow\bm{x}e^{-l} \ , ~~\tau\rightarrow\tau e^{-2l} \ ,
 ~~\Psi_{a,m}\rightarrow\Psi_{a,m}e^{dl/2} \ , \label{bfrrg1}
\end{equation}
at this point when we take $g_0=0=u_{\alpha}$ with $\alpha
=1,2,3$. We will employ this Gaussian fixed point as a point of
departure to study the low-energy physics of $S$ in the dilute
limit and in the unitary regime. That is, we will work under the
assumption
\begin{equation}
 |\mu|, |\delta|\ll \frac{1}{ma_0^2} \ , \label{feshdilute1}
\end{equation}
where $a_0$ is the background scattering length for atoms. We will
take $a_0$ as the short-distance cutoff for the RG flow.

Now we would like to calculate the RG equations at $T=0$. Under
the scaling transformation (\ref{bfrrg1}), $\mu$, $\mu_m$, $g_0$,
and $u_{\alpha}$ transform like
\begin{eqnarray*}
 \mu^{\prime}=\mu e^{2l} \ , ~~\mu_m^{\prime}=\mu_me^{2l} \ ,
 ~~g_0^{\prime}=g_0e^{\epsilon l/2} \ ,
\end{eqnarray*}
and
\begin{eqnarray*}
 u_{\alpha}^{\prime}=u_{\alpha}e^{(\epsilon -2)l} \ ,
\end{eqnarray*}
where $\epsilon =4-d$. ($\delta_c=0$ at the tree level.) Hence,
the coupling constants $u_{\alpha}$'s are irrelevant as long as
$d>2$. However, to guarantee that the Hamiltonian of this system
is bounded from below, we cannot simply set $u_{\alpha}=0$. To
stabilize this system, it suffices to keep the $u_1$ and $u_2$
terms. In the following, we will ignore the $u_3$
term\cite{foot1}.

To proceed, we have to calculate the one-loop RG equations for
$\mu$, $\mu_m$, and $g_0$. Within the $\epsilon$ expansion, we
find
\begin{eqnarray}
 \frac{dr_1}{dl} &=& 2r_1 \ , \label{feshbrg1} \\
 \frac{dr_2}{dl} &=& 2r_2+2\lambda^2r_1-\frac{\lambda^2r_2}{(1-r_1)^2}
 \ , \label{feshbrg11} \\
 \frac{d\lambda}{dl} &=& \frac{\epsilon}{2}\lambda -\frac{\lambda^3}
 {2(1-r_1)^2} \ , \label{feshbrg12}
\end{eqnarray}
with the initial values
\begin{eqnarray*}
 & & r_1(0)=2m\mu a_0^2 \ , ~~r_2(0)=2m\mu_ma_0^2 \ , \\
 & & \lambda (0)=\sqrt{2K_d}mg_0a_0^{\epsilon/2} ,
\end{eqnarray*}
where $K_d=2/[(4\pi)^{d/2}\Gamma (d/2)]$. At this order,
$\delta_c$ is
\begin{eqnarray*}
 \delta_c=2mg_0^2 \! \int^{\Lambda}_{\Lambda e^{-l}} \! \! \frac{d^dp}{(2\pi)^d}
 \frac{1}{p^2} \ .
\end{eqnarray*}
Scaling stops when $l=l_*$ where $l_*$ is defined by
$|r_1(l_*)|=1$. From the solution of Eq. (\ref{feshbrg1}), we get
\begin{equation}
 e^{l_*}=\frac{1}{\sqrt{2m|\mu|a_0^2}} \ . \label{feshbrg2}
\end{equation}
The diluteness condition (the first inequality in Eq.
(\ref{feshdilute1})) leads to $e^{l_*}\gg 1$. Since before scaling
stops $|r_1(l)|\ll 1$, Eqs. (\ref{feshbrg11}) and
(\ref{feshbrg12}) can be approximated as
\begin{eqnarray}
 \frac{dr_2}{dl} &=& (2-\lambda^2)r_2+2\lambda^2r_1 \ , \label{feshbrg21}
 \\
 \frac{d\lambda}{dl} &=& \frac{\epsilon}{2}\lambda -\frac{\lambda^3}
 {2} \ . \label{feshbrg22}
\end{eqnarray}
We want to emphasize that his approximation does not change the fixed-point
structure of Eqs. (\ref{feshbrg1}) -- (\ref{feshbrg12}). In the following,
we will work with Eqs. (\ref{feshbrg1}), (\ref{feshbrg21}), and (\ref{feshbrg22}).

For $d<4$, Eqs. (\ref{feshbrg1}), (\ref{feshbrg21}), and
(\ref{feshbrg22}) have two fixed points: the Gaussian fixed point
$(r_{\alpha},\lambda)=(0,0)$ and the interacting zero-density
fixed point $(r_{\alpha},\lambda)=(0,\lambda_*)$ with
$\lambda_*=\sqrt{\epsilon}$. The Gaussian fixed point is IR
unstable, while the interacting zero-density fixed point is IR
stable. Hence, the latter controls the low-energy physics of the
dilute FR Bose gas in the unitary regime.

To relate $\delta$ to the physical detuning $\nu =-1/a_s$, we will
calculate the $T$-matrix at zero density, and the result is
\begin{eqnarray*}
 T^{-1}(k) &=& \frac{k^2-2m (\delta +\delta_c)}{4mg_0^2} \\
 & & +K_dm \! \int^{\Lambda}_0 \! \! dq\frac{q^{d-1}}{q^2-k^2-i0^+} \ .
\end{eqnarray*}
Using $T(0)=4\pi a_s/m$ and the one-loop result for $\delta_c$, we
get
\begin{eqnarray*}
 \frac{m}{4\pi a_s}=-\frac{m\nu}{4\pi}=-\frac{\delta}{2g_0^2} \ .
\end{eqnarray*}
By definition, the low-energy physics of the system with a wide
Feshbach resonance at zero density is parametrized by only one
parameter $a_s$. Hence, for wide Feshbach resonances, we may
simply take $\lambda (0)=\lambda_*$ if we are only interested in
the physics in the unitary regime\cite{foot2}, and thus we obtain
\begin{equation}
 \delta =\frac{\nu}{bma_0} \ , \label{bfras2}
\end{equation}
in $d=3$ ($\epsilon =1$) where $b=4\pi
K_3/\lambda_*^2=2/(\pi\lambda_*^2)$. Since $\lambda (l)=\lambda_*$
for wide Feshbach resonances, Eq. (\ref{feshbrg21}) can now be
solved easily. Using Eq. (\ref{feshbrg2}) to eliminate $l_*$, we
get
\begin{equation}
 r_2(l_*)=2\mbox{sgn}(\mu)-(2ma_0^2)^{\epsilon/2}\delta |\mu|^{1-d/2}
 \ , \label{bfrrg2}
\end{equation}
where $\mbox{sgn}(x)=1,-1$ for $x>0$ and $x<0$, respectively.

\section{The phase diagram at zero temperature}
\label{phase}

The free energy density $f$ is given by
\begin{equation}
 f=\frac{e^{-(d+2)l_*}}{2ma_0^{d+2}}\tilde{f}(l_*) \ ,
 \label{feshbrgf1}
\end{equation}
where $\tilde{f}(l_*)$ is the free energy density for the
renormalized Hamiltonian at $l=l_*$. Since $\lambda
(l_*)=O(\epsilon)$ and $u_{\alpha}(l_*)\approx 0$, we may
calculate $\tilde{f}(l_*)$ in terms of the perturbation theory in
the coupling constants. The leading order results will be given by
the mean-field theory of the renormalized Hamiltonian. Moreover,
the prefactor in front of $\tilde{f}(l_*)$ in Eq.
(\ref{feshbrgf1}) is always positive, the phases of this system
can be determined directly from $\tilde{f}(l_*)$.

\subsection{Stability conditions}

To proceed, we make a change of variables:
\begin{eqnarray*}
 \bm{x}=a_0\tilde{\bm{x}} \ , ~~\tau =2ma_0^2\tilde{\tau} \ , ~~
 \Psi_{a,m}=a_0^{-d/2}\tilde{\Psi}_{a,m} \ ,
\end{eqnarray*}
the action $S$ [Eq. (\ref{feshbact1})] can be written as
\begin{eqnarray*}
 S &=& \! \int^{\tilde{\beta}}_0 \! \! d\tilde{\tau} \! \int \! \!
 d^d\tilde{x}\tilde{\Psi}_a^{\dagger}(\partial_{\tilde{\tau}}-\nabla^2
 -r_1)\tilde{\Psi}_a \\
 & & + \! \int^{\tilde{\beta}}_0 \! \! d\tilde{\tau} \! \int \! \!
 d^d\tilde{x}\tilde{\Psi}_m^{\dagger} \! \left(\partial_{\tilde{\tau}}
 -\frac{\nabla^2}{2}-r_2\right) \! \tilde{\Psi}_m \\
 & & + \! \int^{\tilde{\beta}}_0 \! \! d\tilde{\tau} \int \! \!
 d^d\tilde{x}~U \ ,
\end{eqnarray*}
where $\tilde{\beta}=\beta/(2ma_0^2)$, $s_{1,2}(0)=\frac{K_d}{16}m
u_{1,2}a_0^{2-d}$, and
\begin{eqnarray*}
 U=\sqrt{\frac{2}{K_d}}\lambda \! \left(\tilde{\Psi}^{\dagger}_m
 \tilde{\Psi}_a^2 \! + \! {\mathrm H.c.}\right) \! \! + \! \frac{16s_1}{K_d}
 |\tilde{\Psi}_a|^4 \! + \! \frac{16s_2}{K_d}|\tilde{\Psi}_m|^4 \ .
\end{eqnarray*}
At large values of $|\tilde{\Psi}_{a,m}|$, the renormalized
potential $U$ at $l=l_*$ is of the form
\begin{eqnarray*}
 \frac{16s^*_1}{K_d}|\tilde{\Psi}_a|^4+\frac{16s^*_2}{K_d}|\tilde{\Psi}_m|^4
 \ ,
\end{eqnarray*}
where $s_{\alpha}^*=s_{\alpha}(l_*)$. In order that the
renormalized Hamiltonian is bounded from below, we must require
that $s_{1,2}^*>0$.

The one-loop RG equations for $s_{1,2}$ are given by
\begin{eqnarray}
 \frac{ds_1}{dl} &=& (2-d)s_1 \ , \label{feshbrg3} \\
 \frac{ds_2}{dl} &=& (2-d)s_2-\lambda^4 \ . \label{feshbrg31}
\end{eqnarray}
In Eq. (\ref{feshbrg31}), we have neglected the $r_1$ and $r_2$
dependence, and thus they are valid only in the dilute limit and
in the unitary regime. Moreover, we do not consider the loop
corrections due to the $u_{1,2}$ terms because they are irrelevant
operators. For the wide Feshbach resonances, we may simply set
$\lambda =\lambda_*$, and thus we get
\begin{eqnarray}
 s_1^* &=& \frac{K_dmu_1}{16a_0^{d-2}}~e^{(2-d)l_*} \ , \nonumber \\
 s_2^* &\approx& \frac{K_dmu_2}{16a_0^{d-2}}~e^{(2-d)l_*}
 -\frac{\epsilon^2}{d-2} \ , \label{bfrrg4}
\end{eqnarray}
in the dilute limit.

The stability conditions require that $u_1>0$ and $u_2>U_c$ where
\begin{eqnarray*}
 U_c &=& \frac{16\epsilon^2}{(d-2)K_dm}~(2m|\mu|)^{1-d/2} \ .
\end{eqnarray*}
In other words, $u_2$ must be strong enough to guarantee the
global stability of this system in the thermodynamical limit. For
given $u_2>0$, the constraint $u_2>U_c$ can also be expressed as a
lower bound for $|\mu|$, i.e.
\begin{equation}
 2m|\mu|a_0^2> \! \left[\frac{16\epsilon^2a_0^{d-2}}{(d-2)K_dmu_2}
 \right]^{2/(d-2)} . \label{bfrrg3}
\end{equation}
Equation (\ref{bfrrg3}) suggests that the FR Bose gas is stable
either at moderate density or with strong dimer-dimer repulsions.

The point that the molecule-molecule repulsion is crucial to
stabilize this system was also emphasized by the previous
mean-field stability analysis\cite{BM}. In the $\mu,\delta <0$
region, the mean-field theory predicts that the ASF state is
stable for $u_2>0$\cite{BM}. Our RG analysis, however, indicates
that in this region this system is thermodynamically unstable as
$|\mu|$ smaller than some critical value. We notice that this
conclusion is consistent with an earlier RPA calculation\cite{Y}.
Within our approach, the reason behind this instability is that
the molecule-molecule interaction becomes attractive at low energy
even when the bare value of $u_2$ is positive.

\subsection{The mean-field theory at $l=l_*$}

To calculate $\tilde{f}(l_*)$, we perform a mean-field theory on
the action $S$ at $l=l_*$. By inserting the mean-field ansatz
$\tilde{\Psi}_{a,m}(\tilde{\tau},\tilde{\bm{x}})=\Phi_{a,m}$ into
$S$ and then taking the variations with respect to
$\Phi_{a,m}^{\dagger}$, we obtain the mean-field equations
\begin{eqnarray}
 -r_1^*\Phi_a+2\sqrt{\frac{2}{K_d}}\lambda_*\Phi_a^{\dagger}\Phi_m
 +\frac{32s_1^*}{K_d}|\Phi_a|^2\Phi_a &=& 0 \ , \nonumber \\
 -r_2^*\Phi_m+\sqrt{\frac{2}{K_d}}\lambda_*\Phi_a^2+\frac{32s_2^*}{K_d}
 |\Phi_m|^2\Phi_m &=& 0 \ , \label{bfrmfe1}
\end{eqnarray}
where $r_{1,2}^*=r_{1,2}(l_*)$. We will solve Eq. (\ref{bfrmfe1})
within the $\epsilon$ expansion. From Eq. (\ref{bfrrg4}), we may
neglect the $s_1^*$ term and the $s_2^*$ term is necessary only
for the existence of the MSF state. Hence, there are three types
of solutions for Eq. (\ref{bfrmfe1}): the ASF state characterized
by $|\Phi_a|=\sqrt{K_dr_1^*r_2^*}/(2\lambda_*)$ and
$|\Phi_m|=\sqrt{K_d/8}r_1^*/\lambda_*$, the MSF state
characterized by $\Phi_a=0$ and
$|\Phi_m|=\sqrt{K_dr_2^*/(32s_2^*)}$, and the vacuum state
(denoted by N) characterized by $\Phi_a=0=\Phi_m$. The solution
for the ASF state can exist only for $r_{1,2}^*>0$, while the
solution for the MSF state can exist only for $r_2^*>0$. In terms
of these solutions, we find that $\tilde{f}_N(l_*)=0$ and
\begin{equation}
 \tilde{f}_{ASF}(l_*)=-\frac{K_dr^*_2}{8\epsilon} \ , ~~
 \tilde{f}_{MSF}(l_*)=-\frac{K_d(r_2^*)^2}{64s_2^*} \ . \label{bfrmfe2}
\end{equation}

The phase diagram can be determined by comparing $\tilde{f}(l_*)$
for the ASF, MSF, and N states. A straightforward calculation
gives rise to the following results: (i) The ASF state occupies
the region with $r_1^*>0$ and $0<r_2^*<8s_2^*/\epsilon$; (ii) the
MSF state occupies the region with  $r_2^*>8s_2^*/\epsilon>0$; and
(iii) the rest in the $r_1^*$-$r_2^*$ plane is the N state.
Substituting the expressions for $r_2^*$ and $s_2^*$ [Eqs.
(\ref{bfrrg2}) and (\ref{bfrrg4})] into result (i), (ii), and (iii),
and taking into account the stability condition [Eq.
(\ref{bfrrg3})], we obtain the phase diagram at $T=0$. The ASF phase
exists in the region with
\begin{equation}
 2m\mu a_0^2> \! \left[\frac{16\epsilon^2}{(d-2)K_dmu_2a_0^{2-d}}
 \right]^{\frac{2}{d-2}} , \label{asfst1}
\end{equation}
and
\begin{equation}
 [2-c_d(2m\mu a_0^2)](2m\mu a_0^2)^{\frac{d}{2}-1}<2m\delta a_0^2<2(2m\mu a_0^2)^{\frac{d}{2}-1}
 \ . \label{asfst11}
\end{equation}
The MSF phase exists in the region with
\begin{equation}
 2m\delta a_0^2<[2-c_d(2m\mu a_0^2)](2m\mu a_0^2)^{d/2-1} \ ,
 \label{msfst1}
\end{equation}
for $2m\mu a_0^2> \!
\left[\frac{16\epsilon^2}{(d-2)K_dmu_2a_0^{2-d}}\right]^{\frac{2}{d-2}}$
and
\begin{equation}
 2m\delta a_0^2<-2(2m|\mu|a_0^2)^{d/2-1} \ , \label{msfst11}
\end{equation}
for $2m\mu a_0^2<- \!
\left[\frac{16\epsilon^2}{(d-2)K_dmu_2a_0^{2-d}}\right]^{\frac{2}{d-2}}$.
Here
\begin{eqnarray*}
 c_d(x)=\frac{K_dmu_2a_0^{2-d}}{2\epsilon}|x|^{d/2-1}-\frac{8\epsilon}{d-2}
 \ .
\end{eqnarray*}
The rest of the phase diagram is occupied by the vacuum state.

\begin{figure}
\begin{center}
 \includegraphics[width=0.9\columnwidth]{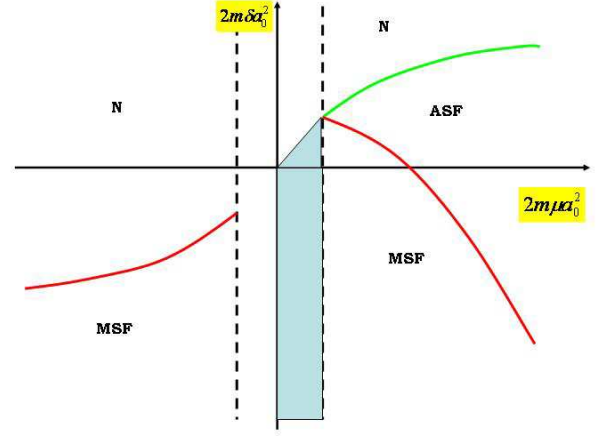}
 \caption{(color online) The schematic phase diagram of a dilute FR Bose
 gas in $d=3$. The unstable region lies between the two dashed lines. In
 the shaded region, the ASF state is an unstable local minimum of the
 free energy density.}
 \label{bfp3}
\end{center}
\end{figure}

A schematic phase diagram is shown in Fig. \ref{bfp3}. The MSF
phase can exist only at finite density of atoms. Furthermore, its
thermodynamical properties do not exhibit universal scaling forms
even in the unitary limit ($\delta =0$) due to the presence of a
dangerously irrelevant operator $u_2$. This can be seen from
$\tilde{f}_{MSF}(l_*)$ [Eq. (\ref{bfrmfe2})]. Although the
thermodynamical instability in the region with $\mu >0$ and
$2m\delta a_0^2<2(2m\mu a_0^2)^{d/2-1}$ (the shaded region in Fig.
\ref{bfp3}) implies that the ASF state will eventually decay, the
time scale, which is governed by the kinetics and dissipation, may
be long enough that the ASF state appears stable. The detailed
mechanism for the decay of the ASF state, however, is beyond the
present study. It is hoped that the ASF state can survive for a
finite lifetime if we prepare the system on the side with a
positive scattering length. (After all, the experiments on
ultracold atoms involve states which are thermodynamically
unstable because the ground state of alkali atoms at nano-Kelvin
temperatures is in fact a solid.) Finally, we must emphasize that
the exact location of the unstable region (the region between the
dashed lines in Fig. \ref{bfp3}) is sensitive to the
short-distance physics, and the determination of it is beyond the
scope of the present approach.

\section{The equilibrium properties at zero temperature}
\label{asf}

From the above analysis, the dilute FR Bose gas may be described
by the ASF state on the side with a positive scattering length
though the ASF state will eventually collapse. Hence, we will
explore the thermodynamics of the ASF at $T=0$.

Inserting $\tilde{f}_{ASF}(l_*)$ [Eq. (\ref{bfrmfe2})] into Eq.
(\ref{feshbrgf1}) and using Eq. (\ref{bfrrg2}), we get the free
energy density for the ASF state
\begin{equation}
 f_{ASF}=-\frac{K_d}{4\epsilon}(2m)^{2-\frac{\epsilon}{2}}
 \mu^{3-\frac{\epsilon}{2}}+\frac{K_d}{8\epsilon}a_0^{\epsilon}
 \delta (2m\mu)^2 \ . \label{bff2}
\end{equation}
From the thermodynamical relation $n=-(\partial f/\partial\mu)$,
we find that
\begin{equation}
 n={\mathcal C}_d(2m\mu)^{\frac{d}{2}} \! \! \left\{1-\frac{1}
 {3-\epsilon/2} \! \left[\frac{2m\delta a_0^{\epsilon}}
 {(2m\mu)^{\frac{d}{2}-1}}\right] \! \right\} , \label{asfmu1}
\end{equation}
where $n$ is the density of (bare) atoms and
\begin{eqnarray*}
 {\mathcal C}_d=\frac{K_d(3-\epsilon/2)}{4\epsilon} \ .
\end{eqnarray*}
Using Euler's relation $f=-P$, we get the equation of state at
$T=0$ for the ASF phase
\begin{equation}
 P=\frac{K_d}{4\epsilon}(2m)^{d/2}\mu^{1+d/2} \! \! \left[1
 -\frac{m\delta a_0^{\epsilon}}{(2m\mu)^{d/2-1}}\right] ,
 \label{asfp1}
\end{equation}
with $\mu$ as a function of $n$ and $\delta$ given by Eq.
(\ref{asfmu1}). Finally, we may calculate the sound velocity $v_s$
through the relation $v_s^2=(n/m)\partial\mu/\partial n$, and
obtain
\begin{equation}
 v_s^2=\frac{2\mu}{dm} \! \left[\frac{(3-\epsilon/2)(2m\mu)^{d/2-1}
 -2m\delta a_0^{\epsilon}}{(3-\epsilon/2)(2m\mu)^{d/2-1}
 -4m\delta a_0^{\epsilon}/d}\right] , \label{bfv1}
\end{equation}
with $\mu$ as a function of $n$ and $\delta$ given by Eq.
(\ref{asfmu1}).

\begin{figure}
\begin{center}
 \includegraphics[width=0.9\columnwidth]{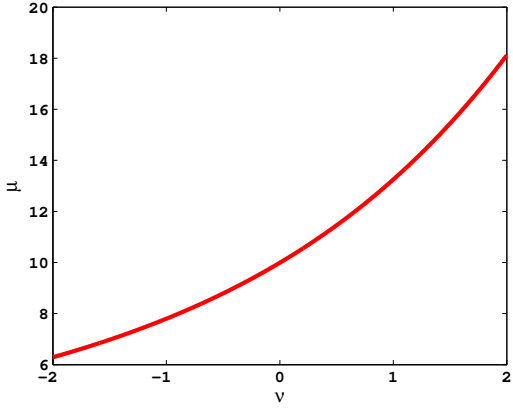}
 \caption{(color online) The chemical potential $\mu$ (in units of
 $n^{2/3}/(2m)$) as a function of $\nu =-1/a_s$ (in units of $n^{1/3}$)
 for given atom density $n$.}
 \label{asfmu3}
\end{center}
\end{figure}

\begin{figure}
\begin{center}
 \includegraphics[width=0.9\columnwidth]{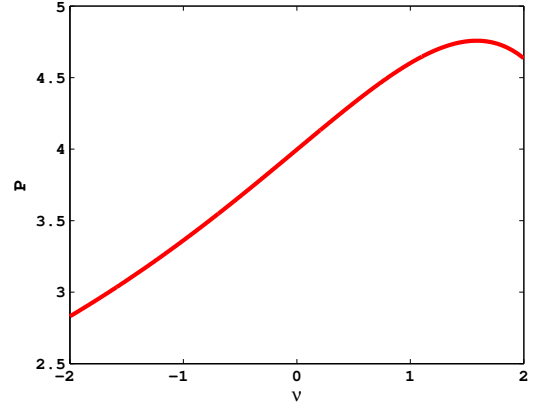}
 \caption{(color online) The pressure $P$ (in units of $n^{5/3}/(2m)$) as
 a function of $\nu =-1/a_s$ (in units of $n^{1/3}$) for given atom density
 $n$.}
 \label{asfp3}
\end{center}
\end{figure}

\begin{figure}
\begin{center}
 \includegraphics[width=0.9\columnwidth]{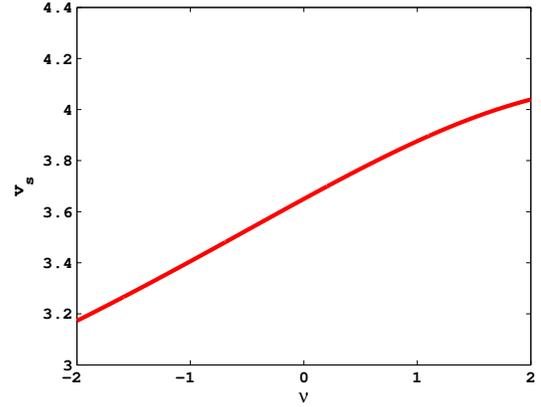}
 \caption{(color online) The sound velocity $v_s$ (in units of $n^{1/3}/(2m)$)
 as a function of $\nu =-1/a_s$ (in units of $n^{1/3}$) for given atom
 density $n$.}
 \label{asfv3}
\end{center}
\end{figure}

The atom density $n$ gives rise to a characteristic length scale
$n^{-1/d}$ and a characteristic energy scale $n^{2/d}/(2m)$. By
measuring the lengths and energies in units of these
characteristic scales, setting $d=3$, and using Eq. (\ref{bfras2})
to replace $\delta$ by $\nu$, Eqs. (\ref{asfmu1}), (\ref{asfp1}),
and (\ref{bfv1}) can be written as
\begin{equation}
 \frac{16\pi^2}{5}=y^{3/2}-\frac{2\pi}{5}xy \ , \label{asfmu4}
\end{equation}
where $x=bm\delta a_0n^{-1/3}=n^{-1/3}\nu$ with $\nu =-1/a_s$,
$b=2/\pi$ and $y=2m\mu n^{-2/3}$, and
\begin{eqnarray}
 \tilde{P} &=& \frac{1}{8\pi^2} \! \left(y^{5/2}-\frac{\pi}{2}xy^2\right)
 , \label{asfp4} \\
 \tilde{v}_s^2 &=& 4y \! \left(\frac{5\sqrt{y}-2\pi x}{15\sqrt{y}-4\pi x}
 \right) , \label{asfv4}
\end{eqnarray}
where $\tilde{P}=2mn^{-5/3}P$ and $\tilde{v}_s=2mv_sn^{-1/3}$.
Equations (\ref{asfmu4}) --- (\ref{asfv4}) exhibit the universal
scaling forms for $\mu$, $P$ and $v_s$. These universal scaling
forms are due to the presence of a nontrivial zero-density fixed
point. The results are shown in Figs. \ref{asfmu3}, \ref{asfp3},
and \ref{asfv3}.

In the unitary limit $\delta =0$ Eqs. (\ref{asfmu1}) and
(\ref{asfp1}) reduce to
\begin{eqnarray}
 \mu &=& \frac{1}{2m} \! \left(\frac{n}{{\mathcal C_d}}\right)^{2/d}
 , \label{asfmu2} \\
 P &=& \frac{n^{1+2/d}}{2m(d/2+1){\mathcal C}_d^{2/d}} \ . \label{asfp2}
\end{eqnarray}
We notice that Eqs. (\ref{asfmu2}) and (\ref{asfp2}) are identical
to the Fermi energy and the equation of state at $T=0$ for a free
Fermi gas with mass $m^*=(d{\mathcal C}_d/K_d)^{2/d}m$. That is,
within the $\epsilon$ expansion, we have shown that the
thermodynamics at $T=0$ for a unitary dilute Bose gas is identical
to the one for a free Fermi gas. This correspondence between the
thermodynamics at $T=0$ for the dilute FR Bose gas and the free
Fermi gas was also noticed in Refs. \onlinecite{wf} and
\onlinecite{SZ} from an entirely different approach. Therefore,
such a ``fermionization" phenomenon should not be an artifact of
either approach, and deserves further investigations.

\section{Conclusions and discussions}

We study the ground-state properties of a dilute FR Bose gas from
the point of view that this system can be described by a
nontrivial zero-density fixed point. From a RG analysis, we first
show that the FR Bose gas will always collapse in the dilute
limit. Following from this consequence, the MSF state cannot exist
in the dilute limit while the ASF state can survive with a finite
lifetime if we prepare this system from the side with a positive
scattering length. Hence, the low-energy properties of a dilute FR
Base gas in the unitary regime are supposed to be described by the
ASF state. Based on this observation, we show that the equilibrium
properties of a dilute FR Bose gas at $T=0$ exhibit universal
scaling forms near the Feshbach resonance.

Most of previous works were concentrated on the mean-field phase
diagram. In the present paper, we focus on the possible universal
properties of this system near the unitary limit. This point was
less studied before, and our RG analysis based on the $\epsilon$
expansion may shed light on this aspect. In particular, unlike
some of previous works\cite{SZ,HO}, the present approach does not
rely on the scaling hypothesis. In stead, the universality in the
unitary regime naturally arises from the presence of a nontrivial
zero-density fixed point. Moreover, the roles of dangerously
irrelevant operators, like the $u_{\alpha}$ terms in the action,
are more easily captured within the RG framework.

A striking result following from our RG analysis is that the $T=0$
thermodynamics of this system in the unitary limit is identical to
that for a free Fermi gas (with a different mass). This point was
also pointed out in Refs. \onlinecite{wf} and \onlinecite{SZ} from
the variational wavefunction approach. Especially, the chemical
potential at zero temperature is of the form $\mu =\alpha
n^{2/3}/m$ in the unitary limit, where $\alpha$ takes a universal
value. The values of $\alpha$ given by Refs. \onlinecite{wf} and
\onlinecite{SZ} are $22.22$ and $6.077$, respectively. Since the
latter is much smaller than the former, the authors of Ref.
\onlinecite{SZ} claimed that the trial wavefunction adopted by
them should be an energetically better candidate for the ground
state. By setting $d=3$ in Eq. (\ref{asfmu2}), we obtain $\alpha
=4.996$. The fact that the obtained numerical value is close to
the one in Ref. \onlinecite{SZ} provides an evidence which further
supports our approximate solutions based on the $\epsilon$ expansion.
However, to determine the exact value of $\alpha$, a more comprehensive
numerical calculation is warranted. Moreover, it deserves further
study whether or not such a correspondence between the dilute FR
Bose gas in the unitary limit and the free Fermi gas can be extended
to finite temperature and non-equilibrium processes.

The quantum phase transition between the ASF and MSF phases
predicted by the mean-field theory\cite{RPW} can be observed only
when these phases are thermodynamically stable. According to the
present analysis (Fig. \ref{bfp3}), the ASF and MSF states can be
stabilized either by strong dimer-dimer repulsions or by
increasing the density of atoms to a moderate value. Therefore, to
observe such a quantum phase transition, we must go beyond the
dilute limit.

Our results are obtained by neglecting the higher-order
correlations such as the three-body effects\cite{3body}.
Therefore, they are applicable before the Efimov physics fully
sets in\cite{Efimov,grimm}. It is interesting to see how the
three-body effects affect the present predictions.

\acknowledgments

The work of Y.-W. Lee is supported by the National Science Council
of Taiwan under grant NSC 96-2112-M-029-006-MY3. The work of Y.L.
Lee is supported by the National Science Council of Taiwan under
grant NSC 98-2112-M-018-003-MY3.



\begin{thebibliography}{99}
 \bibitem{exp1} C.C. Bradley, C.A. Sackett, J.J. Tollett, and R.G. Hulet,
         Phys. Rev. Lett. {\bf 75}, 1687 (1995); C.A. Sackett, J.M. Gerton,
         M. Welling, and R.G. Hulet, {\it ibid.}, {\bf 82}, 876 (1999); E.A.
         Donley, N.R. Claussen, S.L. Cornish, J.L. Roberts, E.A. Cornell,
         and C.E. Wieman, Nature (London) {\bf 412}, 295 (2001); J.L.
         Roberts, N.R. Claussen, S.L. Cornish, E.A. Donley, E.A. Cornell,
         and C.E. Wieman, Phys. Rev. Lett. {\bf 86}, 4211 (2001).
 \bibitem{FKSW} P.O. Fedichev, Yu. Kagan, G.V. Shlyapnikov, and J.T.M.
         Walraven, Phys. Rev. Lett. {\bf 77}, 2913 (1996). 
 \bibitem{exp2} S. Inouye, M.R. Andrews, J. Stenger, H.-J. Miesner, D.M.
         Stamper-Kurn, and W. Ketterle, Nature (London) {\bf 392}, 151
         (1998); S.L. Cornish, N.R. Claussen, J.L. Roberts, E.A. Cornell,
         and C.E. Wieman, Phys. Rev. Lett. {\bf 85}, 1795 (2000); N.R.
         Claussen, E.A. Donley, S.T. Thompson, and C.E. Wieman, {\it ibid.},
         {\bf 89}, 010401 (2002); S.B. Papp, J.M. Pino, R.J. Wild, S. Ronen,
         C.E. Wieman, D.S. Jin, and E.A. Cornell, {\it ibid.}, {\bf 101},
         135301 (2008); S.E. Pollack, D. Dries, M. Junker, Y.P. Chen, T.A.
         Corcovilos, and R.G. Hulet, {\it ibid.}, {\bf 102}, 090402 (2009).
 \bibitem{RPW} L. Radzihovsky, J. Park, and P.B. Weichman, Phys. Rev. Lett.
         {\bf 92}, 160402 (2004); M.W.J. Romans, R.A. Duine, S. Sachdev, and
         H.T.C. Stoof, {\it ibid.}. {\bf 93}, 020405 (2004); L. Radzihovsky,
         P.B. Weichman, and J.I. Park, Ann. Phys. {\bf 323}, 2376 (2008).
 \bibitem{LL} Y.W. Lee and Y.L. Lee, Phys. Rev. B {\bf 70}, 224506 (2004).
 \bibitem{BM} S. Basu and E.J. Mueller, Phys. Rev. A {\bf 78}, 053603 (2008). 
 \bibitem{KMDS} A. Koetsier, P. Massignan, R.A. Duine, and H.T.C. Stoof,
         Phys. Rev. A {\bf 79}, 063609 (2009). 
 \bibitem{wf} S. Cowell, H. Heiselberg, I.E. Mazets, J. Morales, V.R.
         Pandharipande, and C.J. Pethick, Phys. Rev. Lett. {\bf 88}, 210403
         (2002). 
 \bibitem{SZ} J.L. Song and F. Zhou, Phys. Rev. Lett. {\bf 103}, 025302
         (2009). 
 \bibitem{NS} P. Nikoli\'c and S. Sachdev, Phys. Rev. A {\bf 75},
         033608 (2007).
 \bibitem{NS1} Y. Nishida and D.T. Son, Phys. Rev. Lett. {\bf 97}, 050403
         (2006); Y. Nishida and D.T. Son, Phys. Rev. A {\bf 75}, 063617
         (2007); Y. Nishida, {\it ibid.} {\bf 75}, 063618 (2007). See
         also Z. Nussinov and S. Nussinov, {\it ibid.} {\bf 74}, 053622
         (2006).
 \bibitem{VSR} M.Y. Veillette, D.E. Sheehy, and L. Radzihovsky, Phys. Rev.
         A {\bf 75}, 043614 (2007).
 \bibitem{foot1} The inclusion of the $u_3$ term will shift the boundary of
         the unstable region in Fig. \ref{bfp3}, i.e. the value of the R.H.S.
         in Eq. (\ref{bfrrg3}), but does not change the qualitative structure
         of the phase diagram and the results we obtain.
 \bibitem{foot2} For narrow Feshbach resonances, the low-energy physics at
         zero density is parametrized by two parameters: the scattering
         length $a_s$ and the effective range $r_0$. In this case, we cannot
         simply take $\lambda (0)=\lambda_*$. By a similar calculation for
         the $T$-matrix, we may associate $\delta$ and $\lambda (0)$ with
         $a_s$ and $r_0$.
 \bibitem{Y} L. Yin, Phys. Rev. A {\bf 77} 043630 (2008). 
 \bibitem{HO} T.L. Ho, Phys. Rev. Lett. {\bf 92}, 090402 (2004).
 \bibitem{3body} B.D. Esry, C.H. Greene, and J.P. Burke, Jr., Phys. Rev. Lett.
         {\bf 83}, 1751 (1999); E. Braaten, H.-W. Hammer, and T. Mehen,
         {\it ibid.}, {\bf 88}, 040401 (2002). 
 \bibitem{Efimov} V. Efimov, Phys. Lett. {\bf 33B}, 563 (1970); Sov. J. Nucl.
         Phys. {\bf 12}, 589 (1971).
 \bibitem{grimm} T. Kraemer, M. Mark, P. Waldburger, J.G. Danzl, C. Chin, B.
         Engeser, A.D. Lange, K. Pilch, A. Jaakkola, H.-C. N\"agerl, and R.
         Grimm, Nature (London) {\bf 440}, 315 (2006). 
\end{thebibliography}
\end{document}